\documentclass[a4paper]{article}
\usepackage{amsfonts,amssymb,graphicx,latexsym,url,pb-diagram}
\usepackage{amsmath}
\usepackage{fullpage,pict2e,float}
\floatstyle{plain} \newfloat{algorithm}{htb}{}[section]
\floatname{algorithm}{\textbf{Algorithm}}
\usepackage{theorem}
\newtheorem{proposition}{Proposition}[section]
\newtheorem{corollary}{Corollary}[section]
{\theorembodyfont{\rmfamily}
\newtheorem{definition}{Definition}[section]}
{\theorembodyfont{\rmfamily}
\newtheorem{example}{Example}[section]}
{\theorembodyfont{\rmfamily}
}
\usepackage[mathscr]{eucal}
\renewcommand{\theenumi}{\roman{enumi}}

\makeatletter
\renewcommand\p@enumii{(\theenumi)}
\newcommand{\card}{\mathop{\operator@font card}}
\makeatother
\newsavebox{\proofbox}
\savebox{\proofbox}{%
  \begin{picture}(7,7)\put(0,0){\framebox(7,7){}}\end{picture}%
}
\newenvironment{proof}{%
  \list{}{\leftmargin0pt
    \rightmargin\leftmargin}%
  \item[]{\hspace*{1em}\it Proof.\ }%
}
{\hspace*{\fill}{\usebox{\proofbox}}\endlist}
\begin{document}
\title{Fuzzy Chemical Abstract Machines}
\author{Apostolos Syropoulos\\
        Greek Molecular Computing Group\\
        366, 28th October Str.\\
        GR-671\ 00\ \ Xanthi, Greece\\
        \texttt{asyropoulos@yahoo.com}}
\maketitle
\begin{abstract}
Fuzzy set theory opens new vistas in computability theory and here I show this
by defining a new computational metaphor---the fuzzy chemical metaphor. This metaphor 
is an extension of the chemical metaphor. In particular, I introduce the
idea of a state of a system as a solution of fuzzy molecules, that is molecules that 
are not just different but rather similar, that react according to a set of fuzzy
reaction rules. These notions become precise by introducing fuzzy labeled transition
systems. Solutions of fuzzy molecules and fuzzy reaction rules are used to define the 
general notion of a fuzzy chemical abstract machine, which is a {\em realization} of the 
fuzzy chemical metaphor.  Based on the idea that these machines can be used to describe 
the operational semantics of process calculi and algebras that include fuzziness as a 
fundamental property, I present a toy calculus that is a fuzzy equivalent of the
$\pi$-calculus.
\end{abstract}
\section{Introduction}
The Gamma model of parallel programming was introduced by Jean-Pierre Ben\^{a}tre 
and Daniel Le M\'{e}tayer in~\cite{benatre90} (see also~\cite{benatre93} for
a more accessible account of the model and~\cite{benatre01} for a recent account
of it; also see~\cite{calude01} for a thorough presentation of the field of
multiset processing). At the time of its introduction,  parallel
programming as a mental activity was considered more difficult 
(one might also say cumbersome) than sequential programming, something that even today it is 
still valid to a certain degree. Ben\^{a}tre and Le M\'{e}tayer designed Gamma 
in order to ease the design and specification of parallel algorithms. Thus, making
a parallel programming task easier compared to previously available approaches.
Gamma was inspired by the chemical reaction model. According to this metaphor, the 
state of a system is like a chemical solution in which {\em molecules}, that is,
processes, can interact with each other according to {\em reaction rules}, while
all possible contacts are possible since a magic hand stirs the solution continuously. 
In Gamma solutions are represented by {\em multisets} (i.e., an extension of sets 
that is based on the assumption that elements may occur more than one time, 
see~\cite{syropoulos01} for more details). Processes are elements of a multiset and
the reaction rules are multiset rewriting rules. Thus, Gamma is a formalism for 
multiset processing. 
 
The chemical abstract machine (or cham for short) is a model of
concurrent computation developed by G{\'e}rard Berry and G{\'e}rard Boudol~\cite{berry92}. 
The cham is based on the Gamma model and was designed
as a tool for the description of concurrent systems. Basically, each cham is a chemical
solution in which floating molecules can interact with each other according to a set of 
reaction rules. In addition, a magical mechanism stirs the solution so as to allow possible 
contacts between molecules. As is evident, a number of interactions happen concurrently.  

Informally, a process is a program in execution, which is completely characterized by 
the value of the {\em program counter}, the contents of the processor's registers and
the process {\em stack} (see~\cite{silberschatz05} for an overview). Naturally, two
or more processes may be initiated from the same program (e.g., think of a web browser
like FireFox). In general, such processes are considered as distinct execution sequences. 
Although, they  have the same text section, the data sections will vary. Interestingly, one 
can view the totality of processes that run in a system at any moment as a multiset that 
evolves over time, which is built {\em on-the-fly} and may be viewed as the recorded history 
of a system. However, there is a problem that is not covered by this scheme and which I plan 
to tackle here: the processes are not identical, but {\em similar} and so I need a more 
expressive mathematical formalism to describe running processes.

Fuzzy set theory was introduced by Lotfi Asker Zadef in 1965 (see~\cite{klir95} for a thorough
presentation of the theory of fuzzy sets). Fuzzy set theory is based on the denial of the idea that
an element either belongs or does not belong to some set. Instead, it has introduced the notion
of a {\em membership degree} $i$ according to which an element belongs to a fuzzy subset
of an ordinary or {\em crisp} set. Usually, this degree is a real number between zero and one 
(i.e., $0\le i\le1$). For example, I may say that the degree to which $x$ belongs to the fuzzy 
subset $A$ is $0{,}75$. According to fuzzy set theory, when an element belongs to a fuzzy subset
with membership degree equal to zero, this does not imply that the element does not belong to 
a set. On the contrary, it simply means that the element belongs to the fuzzy subset with 
degree that is equal to zero. Formally, given a universe $X$ a fuzzy subset $A$ is characterized 
by a function $A:X\rightarrow[0,1]$, where $A(x)=i$ means that the membership degree of $x$ is 
$i$.

\paragraph{Plan of the paper} I start by giving a ``formal'' definition of chams. Next,
I argue why processes should be viewed as fuzzy entities and then I present a model of fuzzy 
processes that culminates to a brief introduction of fuzzy $X$-machines. The fuzzy chemical
abstract machine is introduced next and this discussion is followed by an exercise in
defining a fuzzy version of the $\pi$-calculus from the description of
a fuzzified chemical abstract machine that originally described the $\pi$-calculus. 
\section{Formal Definition of the Chemical Abstract Machine}
The ingredients of a cham are {\em molecules} $m_1,\ldots, m_n$ that flow in {\em solutions} 
$S_1,\ldots, S_k$. Solutions change according to a number of {\em transformation rules}, which 
determine a {\em transformation relation} $S_{i}\rightarrow S_{j}$. A cham is fully specified by 
defining its molecules, its solutions, and its transformation rules. There are two types of transformation rules: general rules, applicable to any cham, and local rules, which are used 
to specify particular chams. Molecules are terms of algebras associated with specific operations 
for each cham. Solutions are multisets of molecules $[m_k,m_k,\ldots, m_l]$. By applying the 
{\em airlock} constructor``$\lhd$'' to some molecule $m_l$ and a solution $S_k$, one gets the
new molecule $m_{l}\lhd S_{k}$.

A specific rule has the following form
\begin{displaymath}
m_{1}, m_{2},\ldots, m_{k}\rightarrow m_{1}^{\prime}, m_{2}^{\prime},\ldots, m_{l}^{\prime},
\end{displaymath}
where $m_{i}$ and $m^{\prime}_{i}$ are molecules. Also, there are four general transformation
rules:
\begin{enumerate}
\item the {\em reaction} rule where an instance of the right-hand side of a rule replace
the corresponding instance of its left-hand side. In particular, if there is a rule
\begin{displaymath}
m_{1}, m_{2},\ldots, m_{k}\rightarrow m_{1}^{\prime}, m_{2}^{\prime},\ldots, m_{l}^{\prime}
\end{displaymath}
and if the $M_{i}$s and $M^{\prime}_{j}$s are instances of the $m_{i}$s and the 
$m^{\prime}_{j}$s by a common substitution, then
\begin{displaymath}
[M_1, M_2,\ldots, M_{k}]\rightarrow[M_{1}^{\prime}, M_{2}^{\prime},\ldots, M_{l}^{\prime}].
\end{displaymath}
\item The {\em chemical} rule specifies that reactions can be performed freely within any 
solution:
\begin{displaymath}
\frac{S\rightarrow S'}{S\uplus S''\rightarrow S'\uplus S''},
\end{displaymath}
where $\uplus$ is the multiset sum operator.
\item According to the {\em membrane} rule sub-solutions can freely evolve in any context:
\begin{displaymath}
\frac{S\rightarrow S'}{[C(S)]\rightarrow[C(S')]},
\end{displaymath}  
where $C(\;)$ is molecule with a hole $(\;)$ in which another molecule is placed. Note that
solutions are treated as megamolecules.
\item The {\em airlock} rule has the following form:
\begin{displaymath}
[m]\uplus S\leftrightarrow [m \lhd S].
\end{displaymath}
\end{enumerate}
\section{Processes as Fuzzy Multisets}\label{section:ii}
At any moment any operating system is executing a number of different processes.
Quite naturally, many different processes are instances of the same program. For
example, in a multi-user environment, different users may run different instances of
the same web browser. Cases like this can be naturally modeled by {\em multisets}, 
which from a generalization of sets. Any multiset may include more than one copy of
an element. Typically, a multiset $M$ is characterized by a function $M:X\rightarrow\mathbb{N}$,
where $\mathbb{N}$ is the set of natural numbers including zero and $M(x)=n$ means
that multiset $M$ contains $n$ copies of $x$. One may say that the elements of a multiset 
are tokens of different types. Thus, a multiset consisting of two copies of ``a'' and 
three copies of ``b'' can be viewed as  a structure that consists of two tokens of type ``a'' 
and three tokens of type ``b'' (see~\cite{tzouvaras98} for a thorough discussion of this 
idea). When dealing with an operating system, one may argue that types represent
the various programs available to users and tokens represent the various processes
corresponding to these programs.

Although it does make sense to view programs as types and processes as tokens,
still not all tokens are identical. For example, different people that use a particular
web browser view different web pages and have different numbers of tabs open at any 
given moment. Thus, we cannot actually talk about processes that are identical, but we 
can surely talk about processes
that are similar. The next question that needs to be answered is: How can we
compare processes? An easy solution is to define {\em archetypal} processes and
then to compare each process with its corresponding archetypal process. The outcome
of this procedure will be the estimation of a similarity degree. But what exactly is
an archetypal process? Clearly, there is no single answer to this question, but roughly
one could define it as follows:
\begin{definition}
Assume that $p$ is an executable program for some computational platform (operating system,
CPU, etc.), then
an archetypal process $\pi$ of $p$ is the program $p$ running with minimum resources.  
\end{definition}
Here the term {\em minimum resources} means that the archetypal process consumes the least 
possible resources (e.g., memory, CPU time, etc.). 
\begin{example}
Let $f$ be some web browser, then an archetypal process of $f$ is the web browser
which when starts shows a blank page in only one tab/window. Similarly, an archetypal
process for the Unix \texttt{ls} command is the command running without arguments in
an empty directory.
\end{example}
Assume that $p_{1}$ and $p_{2}$ are two processes initiated from the same binary $p$.
Assume also that $\pi$ is an archetypal process and that $\delta_{\pi}$ is a method that 
measures the similarity degree of some process $p_i$ to $\pi$. In different words, 
$\delta_{\pi}(p)=i$ means that  $p$ is similar with $\pi$ to a degree equal to $i$. 
If $\Delta_{\pi}(p_1,p_2)$ denotes degree to which the two processes $p_1$ and $p_2$ 
are similar, then
\begin{displaymath}
\Delta_{\pi}(p_1,p_2)=1-\Bigm|\delta_{\pi}(p_1)-\delta_{\pi}(p_2)\Bigm|.
\end{displaymath}

Suppose that one has selected a number of criteria to choose and specify archetypal
processes. Let $\mathscr{P}_{\sigma}$ denote the set of all possible archetypal processes 
for a particular system $\sigma$ and a particular set of criteria. Without loss of generality, 
we can think that the elements of $\mathscr{P}_{\sigma}$ are the names of all programs that can 
possibly be executed on a system $\sigma$. For instance, for some typical Unix system $S$, the 
set $\mathscr{P}_{S}$ may contain the names of programs under \url{/usr/bin}, \url{/usr/local/bin}, 
\url{/opt/sfw/bin}, etc. Suppose that $\pi\in\mathscr{P}_{\sigma}$ and that at some moment $t$, 
$p_1,\ldots, p_n$ are processes initiated from program $\pi$. Then this situation can naturally
be modelled by fuzzy multisets, that is, multisets whose elements belong to the set to some
degree. A fuzzy multiset is characterized by a higher-order function 
$\Delta:\mathscr{P}_{\sigma} \rightarrow([0,1]\rightarrow\mathbb{N})$. Fuzzy multisets were 
introduced by Ronald R.~Yager~\cite{yager86}. By uncurrying the functional $\Delta$, we get a 
function $\delta:\mathscr{P}_{\sigma}\times[0,1]\rightarrow\mathbb{N}$. Thus, in general, at any 
given moment the processes running in a system can be described by a multiset of pairs 
$(p_{i},\ell_{i})$, where $\ell_i$ denotes the membership degree. However, such a structure
reflects what is happening in a system a given moment. Thus, to describe what is going on
in a system at some time interval we need a structure that can reflect changes as time passes.
The most natural choice that can solve this problem is a form of a {\em set through time\/}.

Bill Lawvere was the first to suggest that sheaves can be viewed as continuously varying sets
(see~\cite{awodey06} and~\cite{bell06} for a detailed account of this idea). Since 
in this particular case I am interested in {\em fuzzy multisets continuously varying through 
time}, it seems that sheaves are not the structures I am looking for. However, as I will
show this is not true. But before I proceed, it is more than necessary to give the general
definition of a sheaf. The definition that follows has been borrowed from~\cite{goldblatt06} 
(readers not familiar with basic topological notions should consult any introductory textbook,
e.g., see~\cite{topology}): 
\begin{definition}
Let $X$ be a topological space and $\mathcal{O}(X)$ its collection of open sets. A sheaf 
over $X$ is a pair $(T,p)$ where $T$ is a topological space and $p:T\rightarrow X$
is a local homeomorphism (i.e., each  $t\in T$ has an open neighborhood $U$ in $T$ that is
mapped homeomorphically by $p$ onto $p(U)=\{p(x)\;|\; x\in U\}$, and the later is open in
$X$).
\end{definition}
Let us now construct a fuzzy multiset through time. Suppose that 
$A:X\rightarrow[0,1]\rightarrow\mathbb{N}$ characterizes some fuzzy multiset. Clearly,
the function $A':X\times[0,1]\rightarrow\mathbb{N}$ also characterizes the same
fuzzy multiset. And if this function characterizes some fuzzy multiset, then it is
absolutely reasonable to say that the graph of this function characterizes the same
fuzzy multiset. Let $M_j$, $j\in J$, where $J$ is a set of indices, be the graphs of all 
functions $A'_{j}$ that characterize fuzzy multisets. In addition, assume that each $M_j$ is an 
open set of a topological space $\mathscr{X}$. Obviously, it is not difficult to define
such a topological space. For example, it is possible to define a metric between points
$((x_k,i_k),n_k)$ and $((x_l,i_l),n_l)$ and from this to define a metric topology. Having
defined a topology on $(X\times[0,1])\times\mathbb{N}$, it is straightforward to define 
a sheaf over $\mathscr{X}$. In particular, if $\mathbb{N}$ denotes the order topology on
the set of natural numbers, then $\mathcal{N}=(\mathbb{N},p)$, where $p:\mathbb{N}\rightarrow
\mathcal{X}$ is a local homeomorphism, is a sheaf over $\mathscr{X}$. In general, such a sheaf 
characterizes a fuzzy multiset through {\em discrete} time. Clearly, this is not the
only sheaf over $\mathscr{X}$ one can define. In fact, one can build a category 
$\mathbf{Sh}(\mathscr{X})$ with objects all sheaves over $\mathscr{X}$ and with arrows 
$k:(A,p)\rightarrow(B,q)$ the continuous maps $k:A\rightarrow B$ such that 
\begin{displaymath}
\begin{diagram}
\node[10]{B}\arrow[8]{se,l}{q}\\[10]
\node{A}\arrow[7]{ne,l}{k}\arrow[15]{e,b}{p}\node[18]{\mathscr{X}}
\end{diagram}
\end{displaymath}  
commutes. In general, the sheaf $\mathcal{R}=(\mathbb{R}^{+}_{0},q)$, where $\mathbb{R}^{+}_{0}$ is
the set of the positive real numbers including zero and here denotes also the order 
topology on this set and $q:\mathbb{R}^{+}_{0}\rightarrow\mathcal{X}$ is a local homeomorphism,
characterizes a fuzzy multiset through {\em continuous} time. It is not difficult to define a monic 
arrow $k:\mathcal{N}\rightarrow\mathcal{R}$ and, thus, to show that $k$ belongs to 
$\mathrm{Sub}(\mathcal{R})$, that is, the collection of sub-objects of $\mathcal{R}$. Thus, we have 
an implicit proof of the following statement:
\begin{corollary}
Sheaf $\mathcal{R}$ contains more information than $\mathcal{N}$.
\end{corollary}
To put it differently, no discrete system can fully simulate a continuous system. 

It should be obvious, that in the most general case one cannot know beforehand all
the components of a sheaf representing a fuzzy multiset through time. When such a structure is 
used to represent processes, then it is {\em noncomputable}, since one cannot construct it using 
some algorithm as it is not possible to foresee what the users will do. 
 
\section{Towards a Fuzzy cham}
A model of computation is a precise description of a conceptual, real, or abstract 
computational device, which can be used to implement certain computational tasks.
Consequently, there are tasks that are implementable and tasks that are not feasible. 
By broadening the notion of computation (e.g., by augmenting the computational machinery
with external agents), it is possible to extend a model of computation. For example, if one
assumes that things happen randomly, then it makes sense to introduce some form of randomness 
in our computation (e.g., think of nondeterministic Turing machines). Thus, it was more than
expected to see the emergence of extended variants of the cham. Indeed, A. Di Pierro, 
C. Hankin, and H. Wiklicky introduced in~\cite{dipiero05} a probabilistic version of cham.
In this model, multisets and mutliset rewriting rules are associated with probabilities so 
as to allow a nondeterministic computation. In general, nondeterministic conceptual computing
devices are not more powerful than their deterministic counterparts. Thus, from the point of 
view of computability theory such models are not that interesting. On the other hand, fuzziness
seems to be a more promising and interesting direction  (e.g., see~\cite{syropoulos06}) and 
so models of computation incorporating fuzziness are more interesting and, at the same time, 
seem to be closer to what happens in reality. 

A cham is defined by specifying the molecule syntax and a set of molecule reaction rules.
The molecule syntax describes some algebra and the transformation rules are actually multiset
rewriting rules that model transitions in a system. Thus, in order to define a fuzzy cham, or
just fucham for simplicity, it is necessary to review known notions from fuzzy algebraic 
theory and to introduce some notion related to fuzzy labeled transition systems.

\paragraph{Fuzzy algebras} Back in 1971, fuzzy groups were introduced by Azriel 
Rosenfled~\cite{rosenfeld71} as a natural extension of groups:
\begin{definition}
Assume that $(X,\ast)$ is a group and that $A:X\rightarrow[0,1]$ is a fuzzy subset of $X$, then 
$A$ is a fuzzy subgroup of $X$ if 
\begin{displaymath}
\min\bigl\{A(a),A(b)\bigr\}\le A(a\ast b^{-1}),\;\mbox{for all $a,b\in X$}.
\end{displaymath}
\end{definition}
By replacing $\min$ with some other fuzzy $t$-norm, one gets a more general structure called a 
$t$-fuzzy group. Also, by following this line of thought it is possible to define more complex 
fuzzy algebraic structures (e.g., fuzzy rings, etc.). But, this is not the only way to fuzzify 
a group. Indeed, M.~Demirci and J.~Recasens proposed in~\cite{demirci04} a fuzzy subgroup to be 
a group $(X,\tilde{\ast})$, where $\tilde{\ast}(a,b,c)$ denotes the degree to which $c=a\ast b$. 
An even more fuzzier algebraic structure is one were both the elements belongs to the 
underlying set to a degree and the result of the operation is the ``real'' result up to some
degree.
\begin{definition}
Assume that $(X,\tilde{\ast})$ is a group and that $A:X\rightarrow[0,1]$ is a fuzzy subset of $X$, then 
$A$ is a fuzzy subgroup of $X$ if
\begin{displaymath}
\min\bigl\{A(a),A(b)\bigr\}\le \tilde{\ast}(a\ast b^{-1},c),\;\mbox{for all $a,b\in X$}.
\end{displaymath}
\end{definition}
Of course when one talks about processes, it does not make sense to speak about real 
algebraic operations. For example, although the notion of the inverse of some program
$P$ has been discussed thoroughly in~\cite{gries81}, there is no general technique to get 
the inverse of some program. Fortunately, one can talk about monoidal operations since 
we can find some identity process. Given two processes $p$ and $q$, the expressions $p+q$ and 
$p\mathbin{|}q$ denote a nondeterministic choice between $p$ and $q$ and the (parallel)
composition of $p$ and $q$, respectively. In particular, $p+q$ behaves either like $p$ or
like $q$. In many case, the environment favors one particular alternative, while in others
it may favor the other alternative. And it is this remark that justifies the introduction of 
fuzzy set theory to model cases like this one. The expression $p\mathbin{|}q$ can be viewed 
as a system in which $p$ and $q$ may proceed independently but may also interact in some 
prescribed way. As time passes by, the system may evolve but its evolution will depend on 
the degree $p$ and/or $q$ can evolve to $p'$ and/or $q'$, respectively. Of course, we can
introduce fuzziness by specifying to what degree two processes may proceed independently.

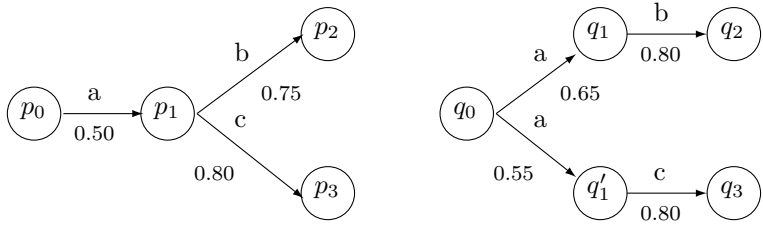
\begin{figure}
\begin{tabular}{lr}
\begin{picture}(150,130)(-50,50)
\put(20,100){\circle{20}}
\put(15,100){$p_0$}
\put(31,100){\vector(1,0){30}}
\put(40,105){a}
\put(35,90){{\footnotesize 0.50}}
\put(70,100){\circle{20}}
\put(65,100){$p_1$}
\put(130,130){\circle{20}}
\put(125,130){$p_2$}
\put(81,100){\vector(4,3){40}}
\put(95,120){b}
\put(105,105){{\footnotesize 0.75}}
\put(130,70){\circle{20}}
\put(125,70){$p_3$}
\put(81,100){\vector(5,-4){40}}
\put(95,95){c}
\put(80,75){{\footnotesize 0.80}}
\end{picture}
&
\begin{picture}(150,130)(-50,50)
\put(20,100){\circle{20}}
\put(15,100){$q_0$}
\put(70,130){\circle{20}}
\put(65,130){$q_1$}
\put(31,100){\vector(4,3){30}}
\put(45,120){a}
\put(55,105){{\footnotesize 0.65}}
\put(70,70){\circle{20}}
\put(65,70){$q'_1$}
\put(31,100){\vector(5,-4){30}}
\put(45,95){a}
\put(30,75){{\footnotesize 0.55}}
\put(120,130){\circle{20}}
\put(115,130){$q_2$}
\put(80,130){\vector(1,0){30}}
\put(90,135){b}
\put(85,120){{\footnotesize 0.80}}
\put(120,70){\circle{20}}
\put(115,70){$q_3$}
\put(80,70){\vector(1,0){30}}
\put(90,75){c}
\put(85,60){{\footnotesize 0.80}}
\end{picture}
\end{tabular}
\caption{Two similar FLTSes.}\label{similar:fltses}
\end{figure}

\paragraph{Fuzzy Labeled Transition Systems} Let us now turn our attention of
fuzzy labeled transition systems. The ideas developed below are 
generalizations of ideas and results presented in~\cite{milner99}. Recall that
a fuzzy binary relation $R$ from a set $S$ to a set $T$ is a fuzzy subset of $S\times T$
(i.e., $R$ is characterized by a function $R:S\times T\rightarrow[0,1]$) and it is
denoted by $R(S,T)$. Similarly, one can define fuzzy $n$-ary relations.    
\begin{definition}
A fuzzy labeled transition system (FLTS) over a crisp set of actions $\mathscr{A}$ is a pair
$(\mathcal{Q},\mathcal{T})$ consisting of 
\begin{itemize}
\item a set $\mathcal{Q}$ of states;
\item a fuzzy relation $\mathcal{T}(\mathcal{Q},\mathscr{A},\mathcal{Q})$ 
called the {\em fuzzy transition relation}.
\end{itemize}
If the membership degree of $(q,\alpha,q')$ is 
$d$,\footnote{One could say that 
the membership degree of a tuple $(q,\alpha,q')$ ``indicates the strength of membership 
within the relation''~\cite[p.~120]{klir95}.} we write 
$q\overset{\alpha}{\underset{d}{\rightarrow}}q'$ to denote that the plausibility degree
to go from state $q$ to state $q'$ by action $\alpha$ is $d$. More generally, if
$q\overset{\alpha_1}{\underset{d_1}{\rightarrow}}q_1 
\overset{\alpha_2}{\underset{d_2}{\rightarrow}}q_2\cdots
\overset{\alpha_n}{\underset{d_n}{\rightarrow}}q_n$, then  $q_n$ is called a {\em derivative}
of $q$ with plausibility degree equal to $\min\{d_1,d_2,\ldots,d_n\}$.
\end{definition}
As in the crisp case, it is quite reasonable to ask when two states in a FLTS should be 
considered equivalent to some degree. For example, what can be said about the two
FLTSes depicted in Figure~\ref{similar:fltses}? In order to be able to answer this
question we need to define the notion of similarity between FLTSes.
\begin{definition}
Assume that $(\mathcal{Q},\mathcal{T})$ is an FLTS and that $\mathcal{S}$ is a fuzzy binary
relation over $\mathcal{Q}$. Then $\mathcal{S}$ is called a {\em strong fuzzy simulation 
over $(\mathcal{Q},\mathcal{T})$ with simulation degree $s$}, denoted by $\mathcal{S}_{(s)}$ 
if, whenever $\mathcal{S}(p,q)\ge s$ if $p\overset{\alpha}{\underset{d_1}{\rightarrow}}p'$, 
then there exists $q'\in\mathcal{Q}$ such that $q\overset{\alpha}{\underset{d_2}{\rightarrow}}q'$, 
$d_2\ge d_1$, and $\mathcal{S}(p',q')\ge \mathcal{S}(p,q)$. 
We say that $q$ {\em strongly fuzzily simulates} $p$ with degree $d\in[0,1]$, 
if there exists a strong fuzzy simulation $\mathcal{S}_{(s)}$ such that $d\ge s$.
\end{definition}
In order to make the notion just defined more clear, I will present an example of 
strong fuzzy simulation.
\begin{example}
Let us consider the two FLTSes depicted in Figure~\ref{similar:fltses} and 
the fuzzy binary relation described by the following table.
\begin{displaymath}
\begin{array}{c|cccc}
  S  & p_0 & p_1 & p_2 & p_3 \\ \hline
q_0  & 0.4 & -   & -   & - \\
q_1  & -   & 0.5 & -   & - \\
q'_1 & -   & 0.5 & -   & - \\
q_2  & -   & -   & 0.6 & - \\
q_3  & -   & -   & -   & 0.6
\end{array}
\end{displaymath}
This fuzzy relation is a strong fuzzy simulation and therefore $p_0$ strongly fuzzily 
simulates $q_0$. To verify this one needs to examine each transition
$q\overset{\alpha}{\underset{d}{\longrightarrow}}q'$ for every pair $S(q,p)>0$ and show
that it is matched by some transition $p\overset{\alpha}{\underset{d'}{\longrightarrow}}p'$.
For example, consider the pair $(q'_1,p_1)$. State $q'_1$ has one transition
$q'_1\overset{c}{\underset{0.80}{\longrightarrow}}q_3$ which is matched by
$p_1\overset{c}{\underset{0.80}{\longrightarrow}}p_3$ because $0.80\ge0.80$ and
$S(q_3,p_3)\ge S(q'_1,p_1)$. Therefore, $q'_1$ strongly fuzzily simulates $p_1$
with degree 0.80.
\end{example}
\begin{definition}\label{sfbs:1}
A fuzzy binary relation $\mathcal{S}(\mathcal{Q},\mathcal{Q})$ is said to
be a strong fuzzy bisimulation over the FLTS $(\mathcal{Q},\mathcal{T})$ with
simulation degree $s$ if both
$\mathcal{S}$ and $\mathcal{S}^{-1}$ (i.e., the inverse of $\mathcal{S}$) are strong 
fuzzy simulations. Any $p$ and $q$ are {\em strongly fuzzily bisimilar with degree} $d$ or 
{\em strongly fuzzily equivalent to degree} $d\in[0,1]$, $p\mathrel{\sim_d}q$, if there is a 
strong fuzzy bisimulation $\mathcal{S}_{(s)}$ such that $d\ge s$.
\end{definition}
Some strong fuzzy bisimulations are constructed by others more simple ones as the proof of
the following propositions shows.
\begin{proposition}\label{sfb:2}
Suppose that each $\mathcal{S}^{(i)}_{(s_i)}$, $i=1,2,\ldots$, is a strong fuzzy bisimulation
with simulation degree $s_i$. Then the following fuzzy relations are all strong 
fuzzy bisimulations:
\begin{displaymath}
\begin{array}{ll}
(1)\; \mathrm{Id}_{\mathcal{Q}}\qquad & (2)\;\mathcal{S}^{-1}_{(s)}\\
(3)\;\mathcal{S}^{(1)}_{(s_1)}\circ\mathcal{S}^{(2)}_{(s_2)} & 
(4)\;\bigcup_{i\in I}\mathcal{S}^{(i)}_{(s_i)}.
\end{array}
\end{displaymath} 
\end{proposition}
\begin{proof}
\begin{enumerate}
\item For the identity relation $\mathrm{Id}_{\mathcal{Q}}(\mathcal{Q},\mathcal{Q})$ it
holds that $\mathrm{Id}_{\mathcal{Q}}(q,q)=1$ and $\mathrm{Id}_{\mathcal{Q}}(q_i,q_j)=0$
for all $q,q_i,q_j\in\mathcal{Q}$ and where $q_i\not= q_j$. In addition, it holds that
$\mathrm{Id}^{-1}_{\mathcal{Q}}(\mathcal{Q},\mathcal{Q})=
\mathrm{Id}_{\mathcal{Q}}(\mathcal{Q},\mathcal{Q})$, which trivially shows that 
$\mathrm{Id}_{\mathcal{Q}}$ is a strong fuzzy bisimulation. 
\item In order to show that 
$\mathcal{S}^{-1}_{(s')}$ is a strong fuzzy bisimulation we need to show that the inverse of 
$\mathcal{S}^{-1}$ (i.e., $\mathcal{S}$) is a strong fuzzy bisimulation, which is obvious 
from definition~\ref{sfbs:1}. 
\item Recall that if $\mathcal{S}^{(1)}(\mathcal{P},\mathcal{Q})$ 
and $\mathcal{S}^{(2)}(\mathcal{Q},\mathcal{R})$ are two strong fuzzy bisimulations with 
similarity degrees $s_1$ and $s_2$, respectively, then $\mathcal{T}(\mathcal{P},\mathcal{R})=
\mathcal{S}^{(1)}_{(s_1)}(\mathcal{P},\mathcal{Q})\circ
\mathcal{S}^{(2)}_{(s_2)}(\mathcal{Q},\mathcal{R})$ is a new fuzzy binary relation such that
\begin{displaymath}
\mathcal{T}(p,r)=\max_{q\in\mathcal{Q}}\min\left[\mathcal{S}^{(1)}_{(s_1)}(p,q),
\mathcal{S}^{(2)}_{(s_2)}(q,r)\right].
\end{displaymath} 
Obviously, $\mathcal{T}(p,r)\ge\min\{s_1,s_2\}$, which shows exactly what we were looking for.
\item Assume that $\mathcal{S}^{(1)}_{(s_1)}(\mathcal{P},\mathcal{Q})$ 
and $\mathcal{S}^{(2)}_{(s_2)}(\mathcal{P},\mathcal{Q})$ are two strong fuzzy bisimulations.
Then their union as fuzzy binary relations is defined as follows.
\begin{displaymath}
\left(\mathcal{S}^{(1)}_{(s_1)}\cup\mathcal{S}^{(2)}_{(s_2)}\right)(p,q)=
\max\left[\mathcal{S}^{(1)}_{(s_1)}(p,q),\mathcal{S}^{(2)}_{(s_2)}(p,q)\right].
\end{displaymath}
From this it not difficult to see that
\begin{displaymath}
\left(\mathcal{S}^{(1)}_{(s_1)}\cup\mathcal{S}^{(2)}_{(s_2)}\right)(p,q)\ge
\min\left[\mathcal{S}^{(1)}_{(s_1)}(p,q),\mathcal{S}^{(2)}_{(s_2)}(p,q)\right],
\end{displaymath}
proves the simplest case. From this it is not difficult to see why the general case holds. 
\end{enumerate}
\end{proof}
The following statement reveals some properties of the strong fuzzy bisimulation.
\begin{proposition}\ \
\begin{enumerate}
\item $\sim_{d}$ is an equivalence relation;
\item $\sim_{d}$ is a strong fuzzy bisimulation.
\end{enumerate}
\end{proposition}
\begin{proof}
\begin{enumerate}
\item 
Recall that a fuzzy binary relation $\mathcal{R}(\mathcal{X},\mathcal{X})$ is a {\em fuzzy equivalence relation} if it is reflexive, symmetric, and transitive.
For reflexivity, it is enough to consider the identity relation 
$\mathrm{Id}_{\mathcal{Q}}(\mathcal{Q},\mathcal{Q})$, which is a strong fuzzy bisimulation.
For symmetry it is enough to say that given a strong fuzzy bisimulation $\mathcal{S}_{(s)}$,
its inverse $\mathcal{S}^{-1}_{(s')}$ is also a strong fuzzy bisimulation. Finally, for
transitivity it is enough for say that the relational composition of two strong fuzzy
bisimulations is also a strong fuzzy bisimulation.
\item This is a direct consequence of Proposition~\ref{sfb:2}.
\end{enumerate}
\end{proof}

\paragraph{Fuzzy $X$-Machines} $X$-machines are a model of computation that has been 
introduced by Samuel Eilenberg~\cite{eilenberg74}. Roughly, given an arbitrary set $X$ 
and a family of relations $\Phi=\{\phi_{i}\}$ where $\phi\subseteq X\times X$, an 
$X$-machine $\mathscr{M}$ of type $\Phi$ is an automaton over the alphabet $\Phi$. Although 
a labeled transition system is not an automaton (e.g., there are no terminal states), still 
it is very easy to define automata using the data of a labeled transition system as a
starting point. Thus, a fuzzy automaton is actually a special case of a fuzzy labeled 
transition system that includes a set of initial states and a set of final states. 

Given a fuzzy automaton over an alphabet (a set) $\Sigma$, a partial function can be defined
\begin{displaymath} 
L_{a}^{-1}(b)=\Bigl\{x\Bigm|x\in\Sigma^{*}, ax=b \Bigr\}   
\end{displaymath}
where $\Sigma^{*}$ is the free monoid with base $\Sigma$ and  
$ax$ is the concatenation of strings $a$ and $x$. Note that 
$L_{a}:\Sigma^{*}\rightarrow\Sigma^{*}$ is the {\em left multiplication} and it is defined
as $L_{a}(b)=ab$. Thus, $L_{a}^{-1}$ is the inverse of the left multiplication. Now, 
by replacing  each edge
\begin{displaymath}
p\overset{\alpha}{\underset{d}{\longrightarrow}}q
\end{displaymath}
of an automaton with an edge
\begin{displaymath}
p\overset{L_{\alpha}^{-1}}{\underset{d}{\longrightarrow}}q,
\end{displaymath}
the result is a new automaton which is a fuzzy $X$-machine. Note that the type of 
such a machine is $\Phi=\{L_{\alpha}^{-1}|\alpha\in\Sigma\}$. Obviously, one can 
construct an $X$-machine even from an FLTS, but the result will not be a {\em machine}
 since it will not have
initial and terminal states. This view is correct when one has in mind the classical view of
a machine as a conceptual device that after processing its input in a {\em finite} number 
of operations terminates. Interestingly, there are exceptions to this view that are widely
used even today. For example, an operating system does not cease to operate and those who 
stop unexpectedly are considered failures. Thus, one can assume that a machine will not terminate
to deliver a result but instead it delivers (partial) results as it proceeds 
(see~\cite{syropoulos08} for more details). On the other hand if states are elements of some
fuzzy subset, then we can say that there is a termination degree associated with each
computation. In other words, a computation may not completely stop or it may stop at
some unexpected moment.  This is a novel idea, since the established view is that a
computation must either stop or it will loop forever. Ideas like this one could be used
to model the case where an external agent abruptly terminates a computation. However, a full 
treatment of these and other similar ideas is out of the scope of this paper.
\section{Fuzzy chams}
Roughly, a fucham can be identified with a solution with fuzzy molecules and a set of fuzzy 
reaction rules.  A solution with fuzzy molecules can be modelled by a fuzzy multiset, 
while fuzzy reaction 
rules can be described by fuzzy transitions. Before presenting a formal definition of fuchams 
let us informally examine whether it makes sense to talk about solutions with fuzzy
molecules and about fuzzy reaction rules. To begin with consider the following concrete chemical 
reaction rule:
\begin{displaymath}
2\mathrm{H}_{2}+\mathrm{O}_{2}\rightarrow 2\mathrm{H}_{2}\mathrm{O}.
\end{displaymath} 
According to the ``traditional'' view, two hydrogen molecules react with one oxygen molecule 
to create two water molecules. A fuzzy version of this reaction rule should involve fuzzy
molecules and it should be associated with a plausibility degree. This is justified by the
{\em fact} that molecules of some chemical element or compound are not identical but rather
similar to a degree with an ideal molecule of this particular element or compound. In
other words, not all hydrogen and oxygen molecules that react to create water are identical.
For example, think of deuterium and tritium as ``hydrogen'' molecules up to some degree
that react with oxygen to produce heavy water, tritiated water and/or partially 
tritiated water, that is, water up to some degree. Thus, the ``water'' molecules produced 
when millions of ``hydrogen'' molecules react with oxygen molecules are not identical but 
just similar (if, in addition, the reader considers hydrogen peroxide, that is, 
$\mathrm{H}_2\mathrm{O}_2$, then things will get really {\em fuzzy}). Obviously, 
the higher the similarity degree, the more likely it is that the reaction will take place. 
And this is the reason one must associate with each reaction rule a plausibility degree. 
Although these ideas may seem unnatural, still  G.F.~Cerofolini and P.~Amato~\cite{amato07} 
have used fuzziness and linear logic to propose an axiomatic theory for general chemistry. 
In particular, they have developed ideas similar to ours in order to explain how chemical 
reaction take place, which means that my proposal, which I call the {\em fuzzy chemical 
metaphor}, is not unnatural at all.

The fuzzy chemical metaphor is essentially identical to the chemical metaphor, nevertheless,
it assumes that molecules of the same kind are similar and not identical. Solutions of fuzzy 
molecules may react according to a number of fuzzy reaction rules, whereas each rule is 
associated with a feasibility degree that specifies how plausible it is that a particular 
reaction will take place. A fucham is an extension of the (crisp) cham that is build around 
the fuzzy chemical metaphor. Like its crisp counterpart, any fucham may have up to four 
different kinds of transformation rules that are described below.   

\paragraph{Fuzzy reaction rules}
Assume that we have a solution with fuzzy molecules that are supposed to react according
to some fuzzy reaction rule. Then the reaction will take place only when the similarity degree
of each molecule is greater or equal to the feasibility of the particular reaction rule.
\begin{definition}\label{feasible:rule}
Assume that $(m_i)_{i=1,\ldots,k}$ and $(m_{j}^{\prime})_{j=1,\ldots,l}$ are archetypal 
molecules. Then
\begin{displaymath}
m_{1},\ldots, m_{k} \overset{}{\underset{\lambda}{\rightarrow}}
m_{1}^{\prime},\ldots, m_{l}^{\prime},
\end{displaymath}
is an {\em ideal} fuzzy reaction rule with feasibility degree $\lambda$ that describes how 
likely it is that molecules $(m_i)$ may react to create molecules $(m_{j}^{\prime})$. Suppose that 
$M_i$ is an instance of $m_{i}$ to degree $\delta_{\pi}(M_i)$, that is, the molecule $M_{i}$ 
is similar to $m_{i}$ with degree equal to $\delta_{\pi}(M_i)$. Then the following fuzzy reaction 
\begin{displaymath}
[M_{1},\ldots, M_{k}] \overset{}{\underset{\lambda}{\rightarrow}}
[M_{1}^{\prime},\ldots, M_{l}^{\prime}]
\end{displaymath}
is {\em feasible} with feasibility degree equal to $\lambda$ if 
\begin{equation}\label{feas:eq}
\min\{\delta_{\pi}(M_1),\ldots,\delta_{\pi}(M_k)\}\ge\lambda.
\end{equation}
The similarity degree of a molecule $M_{j}^{\prime}$ depends on the similarity degrees of the atoms
that make up this particular molecule. The most natural choice is to assume that it is the minimum 
of these degrees.
\end{definition}

It is quite possible to have a situation where the same reacting molecules may be able 
to yield different molecules, something that may depend on certain factors. In different 
words, we may have a solution where a number of different fuzzy reaction rules are {\em potentially
applicable}. In this case, the reaction rule with the highest feasibility degree is {\em really applicable}. 
\begin{definition}
Assume that $S$ is a solution for which the following reaction rules are potentially applicable
\begin{align*}
m_1,\ldots,m_k   & \overset{}{\underset{\lambda_1}{\rightarrow}} 
m'_{1},\ldots,m'_{l_1}\\
m_1,\ldots,m_k   & \overset{}{\underset{\lambda_2}{\rightarrow}} 
m''_{2},\ldots,m''_{l_2}\\
\vdots \\
m_1,\ldots,m_k  & \overset{}{\underset{\lambda_n}{\rightarrow}} 
m^{(n)}_{1},\ldots,m^{(n)}_{l_n}
\end{align*}
and that $\delta_{\pi}(M_{i})$, $i=1,\ldots,k$ are the similarity degrees of the actual molecules 
that are contained in $S$. Then the really applicable rule is the one that satisfies the conditions 
of definition~(\ref{feas:eq}) and whose feasibility degree is the largest among the feasibility
degrees of all potentially applicable rules. 
\end{definition}

\begin{algorithm}
\centerline{\fbox{%
\begin{minipage}{200pt}
\begin{tabbing}
\texttt{0}\=forea\=iff\=iff\=\kill
\>$\$\xi=\min\{M_{1},\ldots,M_{k}\}$ \# similarity degrees\\
\>$\$i=1$\\
\>$\$R=-1$\\
\>$@Rules=()$\\
\>$\$\Lambda=0$\\
\>foreach $\$\lambda$ $(\lambda_1,\ldots,\lambda_n)$ \{\# feasibility degrees\\
\>\> if $(\xi\ge\lambda)$ \{ \\
\>\>\> $\mathrm{push}\; @Rules, [\$\lambda, \$i]$\\
\>\>\> if $(\$\lambda\ge\$\Lambda)$ \{\\
\>\>\>\> $\$\Lambda=\$\lambda;\quad \$R=\$i$\\
\>\>\> \}\\
\>\> \}\\
\>\> $\$i++$\\
\>\}
\end{tabbing}
\end{minipage}
}}
\caption{A Perl pseudo-code that computes the really applicable rule of set of rules.}\label{alg:1}
\end{algorithm}
Using the Perl pseudo-code of Algorithm~\ref{alg:1}, one can compute the really applicable and the 
potentially applicable rules. 

\paragraph{The fuzzy chemical rule}
Mixing up two solutions $S_1$ and $S_2$ yields a new solution $S_3$ that contains
the molecules of both $S_1$ and $S_2$. In other words, $S_3$ is the sum of $S_1$
and $S_2$, or more formally
\begin{displaymath}
S_3=S_1\uplus S_2.
\end{displaymath}
Note that in order to find the sum of two or more fuzzy multisets we work as in the
crisp case (see~\cite{syropoulos01}). Nonetheless, because of restriction~\ref{feas:eq}, the
fuzzy chemical rule takes the following form:
\begin{displaymath}
\frac{\Bigl(S_{1}\overset{}{\underset{\lambda}{\rightarrow}}S_{2}\Bigr)
\quad\Bigl(\forall m_{i}\in S_{3}:\delta_{\pi}(m_{i})\ge\lambda\Bigr)}{%
S_{1}\uplus S_{3}\overset{}{\underset{\lambda}{\rightarrow}} S_{2}\uplus S_{3}}.
\end{displaymath}
Note that this restriction applies to all other general rules.

\paragraph{Fuzzy airlock rule}
The airlock rule creates a new molecule that consists of a solution and a molecule. Therefore, one
needs to define a similarity degree for solutions in order to be able to estimate the similarity
degree of the new molecule. 
\begin{definition}
Suppose that $P$ is a process solution represented by a fuzzy multiset $S$. 
The similarity degree of $P$ to a solution that contains only prototype molecules 
is given by: 
\begin{displaymath}
\Delta_{\delta}(P)=\min\Bigl\{\delta_{\pi}(p)\Bigm| p\in P\Bigr\}.
\end{displaymath}
\end{definition}
If $S$ is a fuzzy solution and $m$ a fuzzy molecule, then
\begin{displaymath}
\frac{\lambda\le\min\Bigl\{\Delta_{\delta}(S),\delta_{\pi}(m)\Bigr\}}{%
[m]\uplus S\overset{}{\underset{\lambda}{\leftrightarrow}} [m\lhd S]}.
\end{displaymath}

\paragraph{Fuzzy membrane rule} Suppose that $\delta_{\pi}(C())$ denotes
the similarity degree of molecule $C()$. Then the fuzzy membrane rule is formulated as follows:
\begin{displaymath}
\frac{\Bigl(S\overset{}{\underset{\lambda}{\rightarrow}}S'\Bigr)\quad
\biggl(\lambda\le\min\Bigl\{\Delta_{\delta}(S),\delta_{\pi}(C())\Bigr\}\biggr)}{%
[C(S)]\overset{}{\underset{\lambda}{\rightarrow}} [C(S')]}.
\end{displaymath}

\section{From machines to calculi: An exercise in reverse ``engineering''}
The cham has been used to describe the operations of various process calculi and 
algebras, which have been proposed to describe concurrency.  Since the cham 
is a special case of the fucham, one can use the fucham to describe the 
operational semantics of any such formalism. Nevertheless, this is almost meaningless as 
there is no reason to use a hammer to hit a nail! On the other hand, it would make 
sense to (try to) describe the operational semantics of a formalism that has been 
designed to describe concurrency in a vague environment. The truth is that there is 
no process algebra or process calculus that is built around the fundamental notion of 
vagueness. Therefore, it is not possible to perform a similar exercise for the fucham. 
A different way to demonstrate its expressive power is to take an existing cham, 
which has been designed to describe some crisp process calculus or algebra, and then to 
try to fuzzify the description and, consequently, to come up with a description of a
(hypothetical?) fuzzy process calculus or algebra. Naturally, this approach does not lead to 
a full-fledged theory of vague or  imprecise concurrency theory, but rather it can be 
considered as an exercise  in defining behaviors from models. For this exercise I will use the 
$\pi$-calculus~\cite{milner99} and the corresponding cham~\cite{berry92}.

The $\pi$-calculus is a mathematical formalism especially designed for the description
of {\em mobile processes}, that is, processes that live in a virtual space of linked
processes with mobile links. The $\pi$-calculus is a basic model of computation that
rests upon the primitive notion of {\em interaction}. It has been argued that interaction 
is more fundamental than reading and writing a storage medium, thus, the $\pi$-calculus
is more fundamental than Turing machines and the $\lambda$-calculus (see~\cite{syropoulos08}
and the references herein for more details). 
  
In the $\pi$-calculus processes are described by process expressions that are defined by the 
following abstract syntax:
\begin{displaymath}
P::= \Sigma_{i\in\mathrm{I}}\pi_{i}.P_{i}\;\Big|\; P_{1}\mathbin{|}P_{2}\;\Big|\;
     \mathrm{new}\;\alpha\;P\;\Big|\;!P\;.
\end{displaymath} 
If $\mathrm{I}=\emptyset$, then $\Sigma_{i\in I}\pi_{i}.P_{i}=\mathbf{0}$ is the 
null process that does nothing. In addition, $\pi_{i}$ denotes an 
{\em action prefix} that represents either sending or receiving a message, or 
making a silent transition:
\begin{displaymath}
\pi::= x(\;y)\;\Big|\; \bar{x}\langle\, y\rangle\;\Big|\;\tau\;.
\end{displaymath}
The expression $\Sigma_{i\in\mathrm{I}}\pi_{i}.P_{i}$ behaves
just like one of the $P_{i}$'s, depending on what messages are communicated to the 
composite process; the expression $P_{1}\mathbin{|}P_{2}$ denotes that both processes
are concurrently active; the expression $\mathrm{new}\;\alpha\;P$ means 
that the use of the message $\alpha$ is restricted to the process $P$; and the 
expression $!P$ means that there are infinitely many concurrently active copies of $P$.

\begin{figure}
\begin{center}
\includegraphics[scale=1]{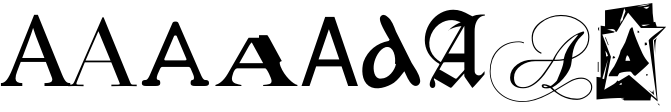}
\end{center}
\caption{Different forms of the same character drawn from different fonts that
demonstrate the notion of typographic similarity.}\label{typo:sim}
\end{figure}
As it stands the only way to introduce fuzziness in the $\pi$-calculus is to assume that
action prefixes are fuzzy. Usually, it is assumed that there is an infinite set of
names $\mathcal{N}$ from which the various names are drawn. In our case, it can be assumed
that names are drawn from $\mathcal{N}\times[0,1]$. In other words, a name would be
a pair $(x,i)$ which will denote that the name used is similar to the prototype $x$ with
degree equal to $i$. Skeptic readers may find this idea strange as an $x$ is always
an $x$ and nothing more or less. Indeed, this is true, nevertheless, if we consider 
various $x$s drawn from different (computer) fonts, then one $x$ is more $x$ than
some other $x$. To fully understand this idea, consider the sequence of letters in
Figure~\ref{typo:sim} borrowed from~\cite{syropoulos08}. Clearly, the rightmost
symbol does not look like an A at least to the degree the second and the third from
the left look like an A. I call this kind of similarity {\em typographic similarity}, 
for obvious reasons. Thus, one can say that names are typographically similar.

Berry and Boudol~\cite{berry92} provide two encodings of the $\pi$-calculus, but for reasons
of simplicity I will consider only one of them. 
The following rules can be used to describe the functionality of the $\pi$-calculus. 
\begin{center}
\begin{tabular}{ll}
$p\mathrel{|}q \rightleftharpoons p,q$ &  (parallel)\\[4pt]
$x(y).p, \overline{x}\langle z\rangle.q \rightarrow p[z/y],q$ &  (reaction)\\[4pt]
$!p \rightleftharpoons p, !p$ & (replication)\\[4pt]
$0\rightharpoonup$ & (inaction cleanup.)\\[4pt]
$\mathrm{new}\;x\;p \rightleftharpoons \mathrm{new}\;y\;p[y/x]$\quad\text{if y is not free in $p$} 
& ($\alpha$-conversion)\\[4pt]
$\mathrm{new}\;x\;p\rightleftharpoons\mathrm{new}\;x\;[p]$ & (restriction membrane)\\[4pt]
$\mathrm{new}\;x\;S, p\rightleftharpoons\mathrm{new}\;x\;[p \lhd S]\quad
\text{if $x$ is not free in $p$}$ & (scope extension)
\end{tabular}
\end{center}
Note that the first four rules are common to the two encodings of Berry and Boudol. Unfortunately,
these rules do not describe summation. However, one can imagine that a sum is an inactive 
megamolecule that changes to a simpler molecule in a single step. Of course, this is a crude idea 
and not a bulletproof solution, so I will not say more on the matter. 

In order to proceed with this exercise, it is necessary to fuzzify the encoding presented
above. Basically, the reaction and $\alpha$-conversion rules are the most problematic
rules. A fuzzification of these rules can be obtained by attaching to each rule a
plausibility degree. In the first case, it is reasonable to demand that the similarity degrees 
of $x$ and $\overline{x}$ are the same and at the same time greater than the feasibility degree 
of the  plausibility degree and also the difference of the similarity degrees of $y$ and $z$ 
is not greater than the plausibility degree. In other words, the reaction 
\begin{displaymath}
x(y).p, \overline{x}\langle z\rangle.q \overset{}{\underset{\lambda}{\rightarrow}} p[z/y],q,
\end{displaymath} 
is feasible if $\delta_{\pi}(x)=\delta_{\pi}(\overline{x})\ge\lambda$ and 
$\delta_{\pi}(z)-\delta_{\pi}(y)\le\lambda$. Similarly, the $\alpha$-conversion
\begin{displaymath}
\mathrm{new}\;x\;p \overset{}{\underset{\lambda}{\rightleftharpoons}} \mathrm{new}\;y\;p[y/x]
\end{displaymath}
if plausible only if $\delta_{\pi}(y)-\delta_{\pi}(x)\le\lambda$. Because of this definition,
it is necessary to define the notion of fuzzy structural congruence. One option is to use
a slightly modified version of the definition provided in~\cite{milner99}. The slight modification
involves $\alpha$-conversion plus $\mathrm{new}\;x\;\mathrm{new}\;y\;P\equiv 
\mathrm{new}\;y\;\mathrm{new}\;x\;P$. From the discussion so far it should be clear that
\begin{displaymath}
\mathrm{new}\;x\;\mathrm{new}\;y\;P\mathrel{\equiv_{\lambda}} 
\mathrm{new}\;y\;\mathrm{new}\;x\;P
\end{displaymath} 
if and only if $\min\{\delta_{\pi}(x),\delta_{\pi}(y)\}\ge\lambda$. With these redefinitions
it is not difficult to go ``back'' to a fuzzy version of the $\pi$-calculus. 
\section{Conclusions}
I have tried to merge fuzziness with concurrency theory. The rationale of this endeavor
is that one can view processes as being similar and not just identical or completely different.
In order to build a model of concurrency in a vague  environment, I have introduced fuzzy
labeled transition systems and proved some important properties. In passing, I have defined
fuzzy $X$-machines and discussed some interesting ideas. The I introduced fuchams as a model 
of concurrent computation in a fuzzy environment where fuzzy processes interact to perform a 
computation according to some fuzzy reaction rules. The model was used to device a toy process
calculus which is a fuzzy extension of the $\pi$-calculus. The next step is to use the
ideas developed here to develop real fuzzy process calculi and algebras and thus to
broaden the study of concurrency. In addition, these ideas may form the basis for developing
fuzzy programming languages and maybe fuzzy computers.

\section*{Acknowledgements}
I would like to thank Athanassios Doupas for his help during the early stages of this work.


\end{document}